\documentclass[aps,prx,twocolumn,amsmath,amssymb,superscriptaddress,groupedaddress]{revtex4}  
\usepackage{graphicx}  
\usepackage{bm}        
\usepackage{amsmath}
\usepackage{amssymb}   
\usepackage{color}
\usepackage{ulem}
\usepackage{multirow}
\usepackage{xcolor}
\usepackage{mathtools}

\begin{document}

\title{Modulation doping and control the carrier concentration in 2-dimensional transition metal dichalcogenides}

\author{N.~Sivadas} 
\email{n.sivadas@samsung.com}
\affiliation{Advanced Materials Lab, Samsung Advanced Institute of Technology-America, Samsung Semiconductor, Inc., Cambridge, Massachusetts 02138, USA}

\author{Y.~Shin}
\email{yongwoo.s@samsung.com}
\affiliation{Advanced Materials Lab, Samsung Advanced Institute of Technology-America, Samsung Semiconductor, Inc., Cambridge, Massachusetts 02138, USA}

\date{\today}

\begin{abstract}
Two-dimensional transition-metal dichalcogenides (TMDs) have attracted interest as post-Si channel candidates in transistor technology. However, despite their potential benefits, controllably doping TMDs has proven difficult. In this work, we proposed a list of candidate elements that can induce p-type and n-type doping in the TMD channel when they are doped onto conventional gate-dielectric oxides.
To verify the screened modulation doping candidates, we demonstrate using first-principles calculations the p-doping of monolayer WSe$_2$ by doping Ni int\textbf{}o the interface dielectric HfO$_2$ layer. 
The induced hole concentration in the WSe$_2$ can be tuned to values compatible with electrostatic gate control of the channel by changing the Ni doping rate. The results of this study will have essential implications for the commercial viability of TMD-based transistors.
\end{abstract}
\maketitle

\section{Introduction}
Over the decades, silicon has been the preferred choice of channel material in transistors due to its reliability, convenience of manufacturing, and natural abundance. However, there are fundamental limits to reducing the dimension of transistors; the performance of Si-based transistors degrades significantly as the channel thickness is reduced below 4~nm~\cite{Schmidt2009, Liu21p43, Su21p2000103}. Reducing the channel layer thickness is crucial for decreasing the channel layer length and avoiding short-channel effects~\cite{Veeraraghavan1989, Young1989}. This has led to a lot of research in finding post-Si channel materials, such as III-V compounds, oxides, and two-dimensional (2D) semiconductors, including TMDs~\cite{ITRS2005, ITRS2015, IRDS23}. Of the many candidate material choices, 2D TMDs have been attracting attention as strong candidates for reducing channel thickness since each layer in TMDs is atomically thin, rendering them the extreme limit of geometric scaling~\cite{Datta2022}. Indeed, recent studies showed that the mobility of monolayer TMDs is comparable with 4 nm thick Si channels with better scaling behavior~\cite{Cui15p534, Wang2016, Ovchinnikov2014, Alharbi16p193502, Allain2014}. 

However, a critical drawback towards the practical application of TMDs is that strategies for controllable doping of TMDs have been largely unsuccessful. Conventional approaches like substitution doping create inhomogeneity, such as charged centers within the channel layer that can severely limit the carrier mobility in the transistor~\cite{Hu2018}. Furthermore, typical dopant centers contribute less than 0.1 carrier to the TMD due to its limited solid-solubility~\cite{Qin2019}, making this method inefficient. Prior research has also explored alternative approaches like modulation doping, introducing organic molecules for doping TMDs~\cite{Le2021, Mouri2013, Kiriya2014}. While using organic molecules modulation doping in TMDs have been demonstrated, using thermally unstable organic molecules in high-volume manufacturing has proven to be a challenge. 

In this paper, we explore the possibility of p- and n-doping TMDs {\it via} modulation doping (see Fig.~\ref{Fig3:Sche}) through an interface of the gate-dielectric oxide layer.  Such oxide-channel interfaces are not only common but also vital in modern field-effect transistors, suggesting that  the proposed modulation doping geometry could be feasible. Using HfO$_2$ as our candidate gate-dielectric layer our first-principles calculations reveals 17 p-doping candidates and 5 n-doping candidates. We also discuss strategies to control the carrier concentration using monolayer WSe$_2$ interfaced with Ni-doped HfO$_2$ as an example; this prescribed approach can be generalized to address materials design challenges where the relative band alignment plays an important role.

\section{Identifying modulation dopants}
\subsection{Toy-model and screening method}

\begin{figure}
\centering
\includegraphics[width=1.0\columnwidth]{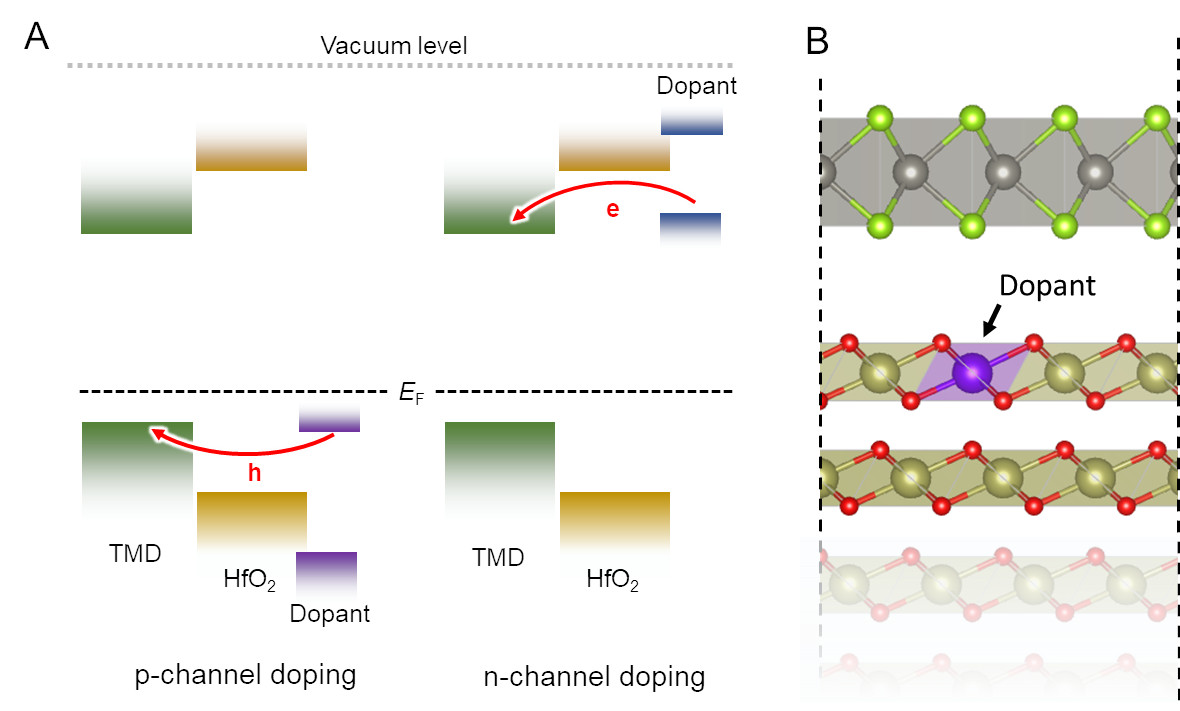}
\caption{(a) Schematic of channel hole (left) and electron (right) doping via modulation doping the interface oxide layer. (b) Crystal structure of monolayer TMD (green) interfaced with doped (violet) HfO$_2$.}
\label{Fig3:Sche}
\end{figure}

Materials with type-III band alignment to TMDs can dope the TMD layer; such band alignment can be induced by modulation doping. We started with a simple model to understand modulation doping and screening the candidates. 
When the conduction band minima (CBM) of the candidate material is lower than the valance band maxima (VBM) of the TMD layer, electrons can be transferred from the TMD layer to the oxide to lower the total energy (see Fig.~\ref{Fig3:Sche} (a) left). This leads to a p-doping of the TMD layer. Similarly, when the VBM of the candidate material layer is higher than the CBM of the TMD layer, this results in n-doping of the TMD layer (see Fig.~\ref{Fig3:Sche} (a) right). 

In all modern transistor architectures, the interface between the channel material and a high-$\kappa$ gate dielectrics layer such as HfO$_2$ is ubiquitous, which provides equivalent scaling~\cite{Datta2022}.
Consequently, for experimentally relevant investigation guided by the above model, we focus our search on studying the effect of substitution doping on the high-$\kappa$ gate dielectric layer on the TMD bands, using HfO$_2$ as a representative gate dielectric oxide. 

The basic concept of our approach is similar to conventional substitution-induced carrier doping. Yet, the critical distinction is that the dopants are not introduced in the channel layer but in the gate oxide in contact with the channel. Such an approach can be expected to minimize carrier scattering from the charged defect centers within the channel layer~\cite{Lee2019}, thereby enhancing the device's performance. Furthermore, as the dopant atom is not directly introduced, the channel layer can be grown under optimal conditions to minimize impurities, one of the biggest challenges in 2D materials growth \cite{Zhao2023}. However, the doping concentration should be carefully selected to not hamper the high-$\kappa$ behavior of the gate dielectrics, which will be discussed in Chapter 2.

To precisely study the doping effect of the 2D channel interface with gate oxide, we simulated a hetero-structure geometry with monolayer TMD interfaced with 15 layers of cubic HfO$_2$ (see Fig.~\ref{Fig3:Sche} (b)). The difference between the in-plane lattice constants of TMD and HfO$_2$ introduces a nominal strain at the interface (5$\%$ on WSe$_2$-HfO$_2$ interface and 9$\%$ on MoS$_2$-HfO$_2$ interface). The effect of this tensile strain on WSe$_2$ is to reduce the local band gap to 0.5~eV and to introduce indirectness to it, consistent with prior reports~\cite{Postorino2020}~\footnotemark[2]\footnotetext[2]{We also constrained the in-plane lattice constants to that of the WSe$_2$ lattice constants and verified that the key results presented are qualitatively unaffected.}.
Along with this constraint, we replaced the interface Hf atoms with candidate dopant atoms. For the appropriate choice of dopant atoms, electrons can be transferred from the oxide layer to the TMD layer or vice versa. This will result in n-doping or p-doping of the TMD layer characterized by the CBM or the VBM of the TMD layer crossing the Fermi level, respectively. 
Note that although such an interface with one entire HfO$_2$ layer replaced by another oxide layer reflects the ideal delta doping geometry, the delta doping limit is sufficient since studying this extreme limit allows us to give bounds on the maximum charge that can be transferred $via$ modulation doping.

\begin{figure*}
\includegraphics[width=1.0\textwidth]{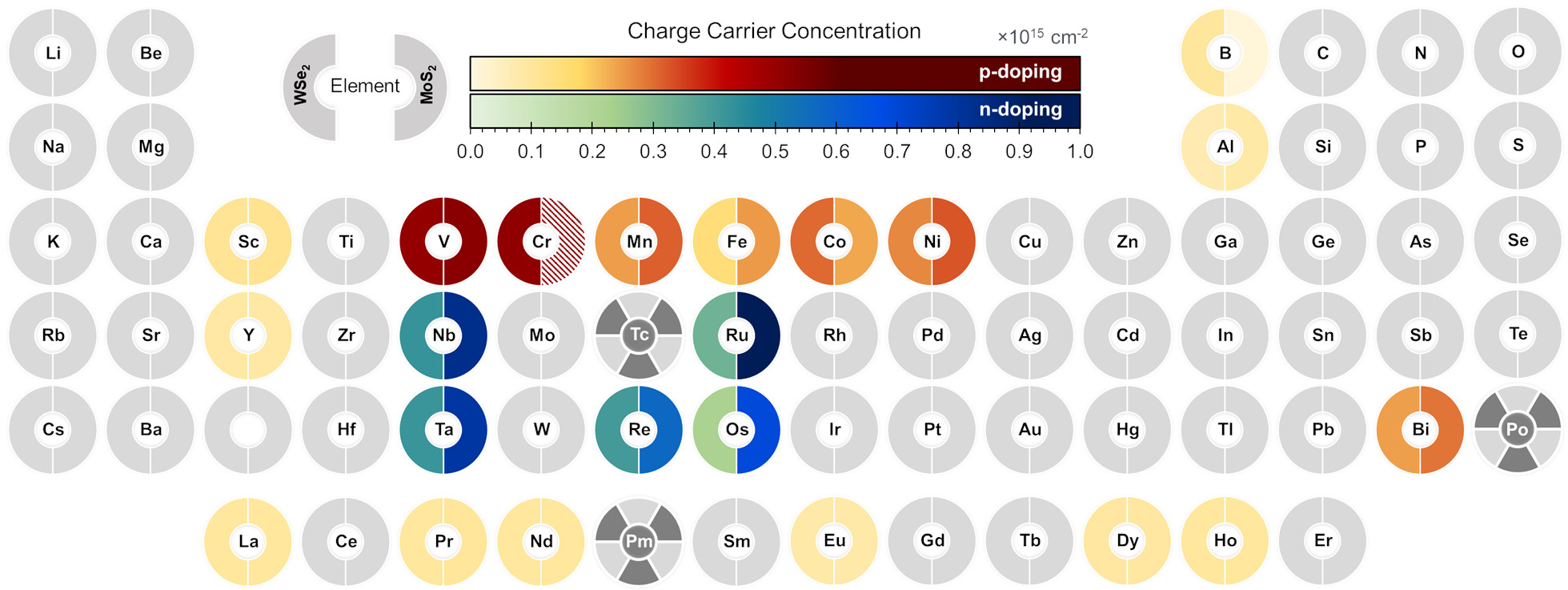}
\caption{\label{Fig2:HTR} Results from our screening process for identifying candidates for p-doping and n-doping TMDs. Radioactive elements are  labeled~\cite{ToxicRadioactive}. Results for WSe$_2$ (left) and MoS$_2$ (right) are shown using heat maps. We identify 17 candidates for p-doping (in red) and 5 candidates for n-doping TMDs (in blue). } 
\end{figure*}

\subsection{Results}\label{KeyRes}

We used a two-step approach to find a set of all candidate dopants to consider to dope the Hf-site in the gate oxide. In the first step, using the atomic layer deposition (ALD) database (DB), we identified 68 unique elements that can be experimentally grown using ALD~\cite{ALD-DB}. Of them, we found 44 cations that form in either 3$^+$, 4$^+$, or 5$^+$ states. In the second step, using the materials project DB~\cite{Ong13p314, Munro20p112}, we removed all ternary Hf-based oxides that has a band gap smaller than 0.1~eV. While this filter can be mitigated depending on the growth details, we use this to avoid dopants from hindering the high-$\kappa$ dielectric behaviors of the HfO$_2$ host material. Screening with these criteria led to 27 elements for further investigation.

The results from our first-principles calculations for both WSe$_2$ and MoS$_2$ are summarized in Fig.~\ref{Fig2:HTR}. The red heat map shows the charge transfer between the interface layer leading to p-type doping of TMDs. Similarly, the blue heat map is used to highlight suitability for n-type doping. The heat maps essentially quantify the extend of the band-overlap and the charge carriers that are transfered to the TMD layer.

While p-type doping is generally considered challenging in most semiconductors, including TMDs,  using our screening procedure we identify 17 candidate dopants that can lead to hole transfer to the TMD layer when doped to the Hf site of gate oxide. This sharply contrasts other approaches, which have largely been unsuccessful in finding good p-doping candidates. We find that V and Cr leads to the most charge transfer in WSe$_2$\footnote{We comment that the absence of a charge transfer in Cr in MoS$_2$ in our findings is an artifact of the lattice constraint relaxation we performed in MoS$_2$ interfaces to maintain the local band-gap which otherwise vanishes from strain effects.}, which is followed by Co, Ni, Mn, and Bi. As V and Bi form stable ternary phases with Hf and O (such as HfV$_2$O$_7$ and HfBiO$_4$, respectively), the dopant atoms in the ternary oxide are unlikely to form at the interface.  This might limit the modulation doping mediated charge transfer as the hybridization of the wavefunctions between gate oxide and TMD can be limited.

Using our model of simple band alignment would lead us to conclude that finding n-doping candidates {\it via} modulation doping of a gate-oxide layer is more challenging than p-doping because oxides typically have a high ionization energy. This limits materials with a type-III band alignment relative to TMDs. While we find fewer candidates to dope the oxide layer in n-doped TMDs than p-doped TMDs, we successfully identify 5 candidates dopant (Nb, Ta, Re, Ru, and Os) when doped into the Hf-site can transfer electrons to the interface TMD layer. Our results highlight the importance of our model going beyond the conventional strategy. Since two of our strong n-doping candidates, Nb and Ta, are considered good candidates for p-doping TMDs when doped directly onto the transition metal site in the TMD layer~\cite{Chua16p5724, Li20p6276, Li22p7662}. Therefore, experimentally observing this n-type modulation doping behavior will require high-quality interfaces where the metal atoms at the oxide interface do not diffuse into the TMD layer. Another solution to overcome this issue is to grow a thin spacer layer, such as h-BN, between the doped oxide layer and the TMD layer to prevent intermixing of the metal atoms at the interface.


Since the list of candidates for n-doping was small, we also extended our search space to substitute anion of gate oxide. We found that halide counterparts of popular gate-dielectric oxides (TiO$_2$, ZrO$_2$, HfO$_2$) of the form MX$_2$ (M = [Ti, Zr, Hf] and X = [Cl, Br, I]) can lead to n-doping of TMDs. This result is consistent with the fact that anions contribute more to the VBM, and replacing the oxygen atoms can modify the ionization energy. However, Cl and Br should be excluded from consideration due to their toxicity. Therefore I-doping of the gate-oxides made up of any combination of TiO$_2$, ZrO$_2$ and HfO$_2$ could induce controllable n-doping in the interface TMD channel layer.

Finally, we want to contrast our results using explicit DFT calculations to high-throughput screening using our model. Using existing DBs like C2DB~\cite{Haastrup2018}, we performed DB screening for materials with type-III band alignment relative to TMDs. One would expect a DB-based search to yield similar results much faster. While there are similarities between the two results, there are notable differences. This further highlights the importance of going beyond a simple band alignment model, which ignores the details of the interface physics. 

\subsection{Discussion towards device application}

As our plan identified elements that can be doped into conventional gate-dielectric oxides like HfO$_2$, this approach is compatible with all the existing transistor geometries, including gate-all-around structure. Further, as the modulation doping mechanism is built around relative band-alignment arguments, we expect the list of dopants identified for HfO$_2$ also to dope other conventional binary gate-dielectric oxides with similar ionization energy and electron affinity (e.g. ZrO$_2$). Our approach should also be agnostic to how the dopant is introduced into the gate oxide. Therefore, we expect our results to be qualitatively valid for both delta-doped and point-doping of the gate oxide.

It is common practice to grow HfO$_2$ in the amorphous form when used as a gate-dielectric to minimize leakage current~\cite{Li2011}. It is worth noting that even though the results discussed are from perfect crystal phases, as long as the dopant atoms are in the appropriate oxidation state and are close to the interface to allow charge transfer, we expect the proposed effect to persist even for amorphous gate-dielectric oxide. 
Indeed, the additional entropy in the amorphous phase could stabilize the dopant atoms in locally unstable atomic environments that can be challenging in their corresponding crystalline phases~\cite{Sun2016}.

\begin{figure}
\centering
\includegraphics[width=1.0\columnwidth]{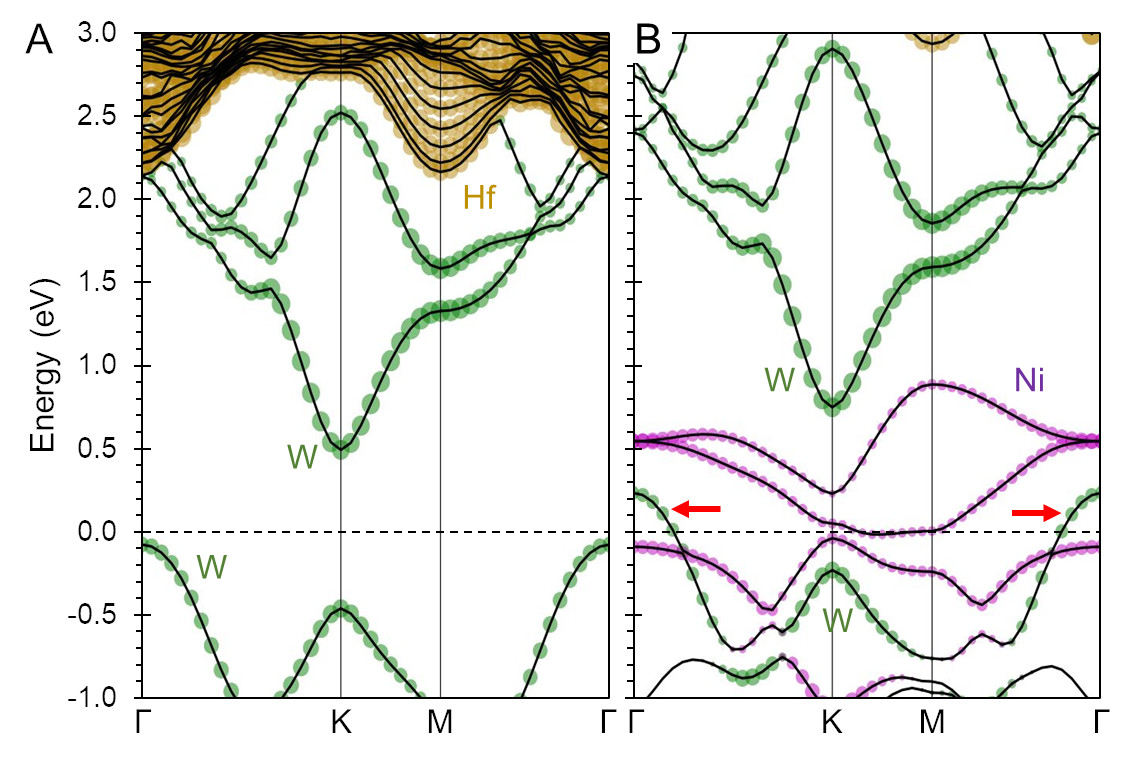}
\caption{\label{Fig4:ES} The atom projected electronic structure of WSe$_2$ (green) interfaced with (a) undoped HfO$_2$ (gold) and (b) Ni-doped HfO$_2$ where the projection onto Ni atoms is shown in magenta. Ni-doped HfO$_2$ leads to p-doping of WSe$_2$ layers (red arrows).}
\end{figure}

\begin{figure*}
\includegraphics[width=1.0\textwidth]{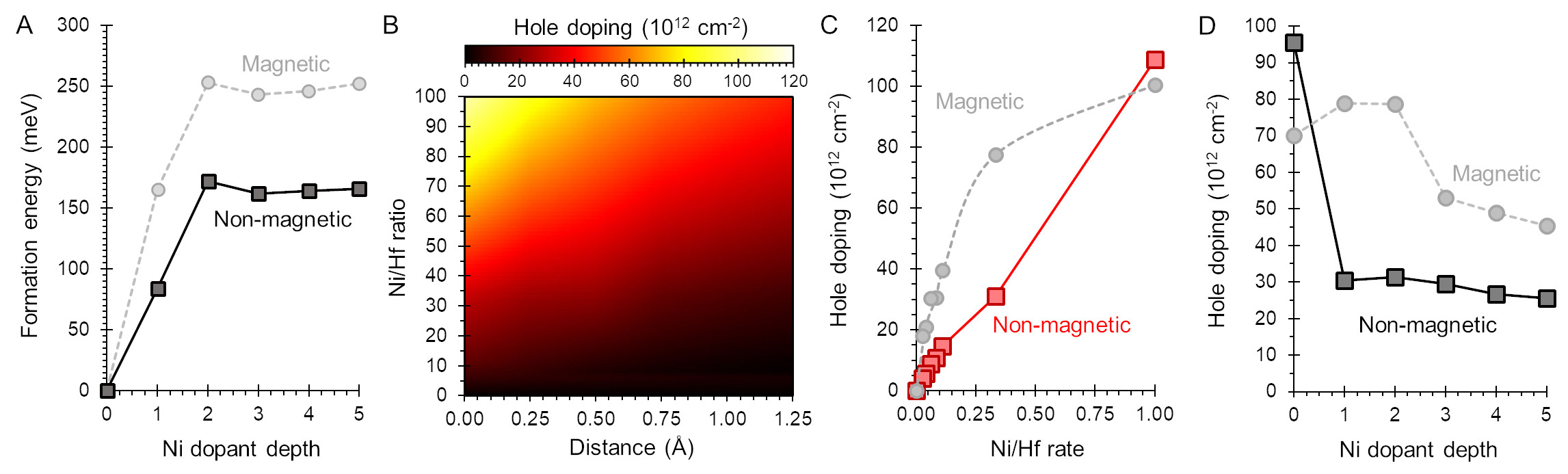}
\caption{\label{Fig5:Control} Hole-doping of WSe$_2$ using Ni-doped HfO$_2$. (a) The formation energy (meV) of Ni doped HfO$_2$ as a function of depth from the interface. (b) The doped hole concentration in WSe$_2$ as a function of the interlayer distance (between the W atom and the interface metal atom) and Ni doping ratio. (c) The hole concentration in WSe$_2$ for an interlayer distance of 4.97~{\AA} for the case of nonmagnetic (in red, same as in (b)) and magnetic cases/ (d) The hole concentration in WSe$_2$ for the realistic interface with one full HfO$_2$ layer replaced by NiO$_2$ as a function of the location of Ni atoms relative to the interface for magnetic and nonmagnetic Ni cases. The hole concentration in WSe$_2$ can be controlled by controlling the Ni doping rate and the positioning of the Ni dopants relative to the interface.}
\end{figure*}

\section{Interface electronic structure and controlling the carrier concentration}
\subsection{Validation of the type-III band alignment}

To check the validity of our screening process, we performed detailed first-principle calculations with Ni, one of the identified p-doping candidates. This choice was motivated by the fact that Ni is cheap (\$14/kg)~\cite{PriceOfElements} and Ni$^{4+}$ state is experimentally stable~\cite{Petousis2017, Zagorac2019}. Since the other experimentally observed competing Ni-oxide phases (NiO, Ni$_2$O$_3$) are non-metallic~\cite{Kedesdy54p5941,Sherman19p584,Jouini19p163109}, some local bonding variations would not lead to a deterioration of the dielectric properties. We considered a heterostructure geometry with Ni-doped HfO$_2$ interfaced with monolayer WSe$_2$. Fig.~\ref{Fig4:ES} contrasts the electronic structure of the undoped interface between HfO$_2$ and WSe$_2$ (a), and the Ni-dope HfO$_2$ interfaced with WSe$_2$ (b). In both calculations, WSe$_2$ interfaces with fifteen layers of gate oxide layers. The pristine (undoped) system has fifteen layers of HfO$_2$, and Ni-doped HfO$_2$ has 100$\%$ of the Hf atoms at the interface layer was replaced by Ni atoms with additional fourteen HfO$_2$ layers, which essentially simulates a delta-doping on gate oxide.

Figure ~\ref{Fig4:ES} shows the atom projected bandstructure along the 2D high-symmetry directions for the undoped HfO$_2$ interfaced with monolayer WSe$_2$ (see Fig. ~\ref{Fig4:ES} (a)) and the Ni-doped HfO$_2$ interfaced with monolayer WSe$_2$ (see Fig. ~\ref{Fig4:ES} (b)). For WSe$_2$ interfaced with undoped HfO$_2$, we observe a type-I band alignment with no charge transfer, as expected. In sharp contrast, WSe$_2$ interfaced with Ni-doped HfO$_2$ shows clear evidence of p-doping with type-III band alignment (red arrows).

Next, to quantify the hole concentration at the interface from the charge transfer of the Ni-doped gate oxide system, we computed the integral of the partial density of states of the W and Se atoms from the Fermi level to the top of the VBM. For the delta-doped case, we found this to be 9.6 × 10$^{13}$ cm$^{-2}$. For practical applications, we estimate that the carrier concentration in the channel layer around the gate dielectric needs to be in the range of 10$^{12}$ - 10$^{13}$ cm$^{-2}$~\cite{Radisavljevic13p815, Tu21p150422}. If the carrier concentration is higher, the excess charge could lead to increased screening of the electric flux lines, rendering the gate control ineffective~\cite{Siao2018}. Accordingly, it is worth systematically investigating the carrier concentration by dopant distributions and rate. Hence, we demonstrate that the hole doping concentration in WSe$_2$ can be controlled by (1) changing the distance between the dopant Ni atoms and TMD layer and (2) tuning the Ni doping rate in HfO$_2$. 

\subsection{Formation energy}

First, we computed the Ni-dopant formation energy as a function of layer depth from the (111) surface of cubic HfO$_2$ without considering the interfacing WSe$_2$ layer. To simulate experimentally relevant growth conditions, we introduced a slab geometry. Nine oxide layers were sufficient to achieve converged results since the defect effect on the fourth layer from the surface is energetically comparable to the bulk\cite{Shin16p38}.  Our calculations at 0~K suggest that Ni prefers to be magnetic. However, at room temperature and under dilute doping Ni can remain nonmagnetic. In Fig.~\ref{Fig5:Control} (a) we report the relative formation energy for Ni doping of HfO$_2$ for both magnetic (grey circles) and nonmagnetic (black rectangles) calculations with the surface layer labeled as `0'. Away from the surface, the relative energies first increase for both magnetic and non-magnetic calculations and then converge as the Ni depth is $\sim$ 10~{\AA} away from the surface. Clearly, Ni doping of HfO$_2$ is relatively stable at the surface compared to the bulk. This is also consistent with the finding that there are no stable ternary phases with Ni, Hf, and O~\cite{Sun2016}.

\subsection{Controlling the carrier concentration by tuning the Ni-doping}

Next, to study the effect of changing the Ni doping rate on the induced hole concentration in WSe$_2$, we used a slab-vacuum model of monolayer WSe$_2$ with monolayer HfO$_2$ to simulate the interface. We created supercells of different sizes where in each case one Hf atom is replaced by a Ni atom to simulate the different Ni doping rates. While such an interface where one layer of WSe$_2$ is interfaced with only one layer of the oxide layer is artificial to compare with experiments, this is a reasonable compromise to overcome the computational complexity of such a simulation. The effect of structural relaxation is also ignored in our calculations. However, we will show below that this interface model gives valuable insights into understanding the effect of varying Ni doping on the induced hole concentration in the WSe$_2$ layer. 
We will also show that the results of such a simple model are consistent with explicitly relaxed experimentally relevant geometries considered in Fig.~\ref{Fig4:ES}. 
Additionally, we also vary the interlayer distance between the WSe$_2$ and the dopant oxide layer (defined as the distance between the W atom and the dopant atom in the oxide layer).

Figure~\ref{Fig5:Control} (b) summarizes our results from the interface model. We see a strong dependence of hole concentration on the Ni doping rate as well as the interlayer distance. A reduction of the Ni doping rate in the HfO$_2$ leads to a reduction in the hole concentration per unit area in the TMD layer. We also see that the hole concentration drops off as the interlayer distance is increased from the optimal distance (4.97~{\AA}; calculated in section 2-A). This is consistent with the expectation that the interlayer interactions are mediated by the hybridization of wavefunctions between layers. As the distance between the W and Ni atoms increases the charge transfer decreases, leading to a smaller hole doping. Thus, by tuning either the Ni doping rate in HfO$_2$ or the distance of the dopant atom from the interface, the induced carrier concentration can be controlled and reduced to a range (10$^{12}$ $\sim$ 10$^{13}$ cm$^{-2}$) where gate-control of carriers is possible.

The controllable hole-doping in WSe$_2$ by tuning the Ni doping rate is more evident in Fig.~\ref{Fig5:Control} (c). It shows the induced hole concentration in WSe$_2$ at an interlayer distance of 4.97~{\AA} when Ni is magnetized (gray circles) and nonmagnetic (red rectangles). This interlayer distance corresponds to the optimal interlayer distance for the previously considered fully relaxed interface of monolayer WSe$_2$ with Ni-doped HfO$_2$ with fifteen oxide sub-layers (see Fig.~\ref{Fig4:ES}). 
Evidently, when the Ni doping rate is reduced, the hole concentration drops systematically. 

Note that for an interlayer distance of 4.97~{\AA} the total hole concentration for the 100$\%$ nonmagnetic-Ni doping case is 10.0 $\times$ 10$^{13}$ cm$^{-2}$. This value is similar to the hole concentration of 9.6 $\times$ 10$^{13}$ cm$^{-2}$ that we reported earlier for the fully relaxed interface of monolayer WSe$_2$ with Ni-doped HfO$_2$ with fifteen oxide sub-layers (optimal distance 4.97~{\AA}). This agreement between the hole concentration predicted using the simplified interface model and the hole concentration predicted from fully relaxed multilayer calculations further validates the predictive capability of our simplified interface model.

For completeness, we computed the effect on the hole concentration on the Ni depth using an interface of monolayer WSe$_2$ and nine sub-layers of oxides. The three sublayers of the oxide farthest away from the interface were constrained to the bulk values to simulate bulk behavior. Due to our computational limitations, we restricted our study to 100 $\%$ Ni doping of a sublayer but varied the Ni doping depth between the six sub-layers near the interface. 

The results are summarized in Fig.~\ref{Fig5:Control} (d) with the dopant located on the interface layer labeled as `0'. For the magnetic case (gray circles), we notice a non-monotonous change where the induced hole concentration first slightly increases away from the interface and then decreases as the Ni atom is further away from the interface. For the nonmagnetic case (black rectangles), we find a sharp drop in the induced hole concentrations as the Ni dopant sublayer is further away from the interface. The reduction of the hole concentration as the Ni atoms are away from the WSe$_2$ layers is consistent with our findings from the interface model (see Fig.~\ref{Fig5:Control} (b)). However, even when the Ni atom is 5 sublayers away from the interface (interlayer distance $\sim$ 20 Å) there is a rather large hole concentration induced in the WSe$_2$ layer. This is in contrast to our findings in Fig.~\ref{Fig5:Control} (b) where the drop in hole concentration with interlayer distance was more prominent. We attribute this discrepancy to the difference in the dielectric environments surrounding the Ni atom in the two cases. Regardless, the overall trend is consistent with our findings from the interface model. We conclude that our argument is valid for any phase of gate-dielectrics, including amorphous.

\section{Conclusion}

In summary, using a combination of high-throughput screening by DFT calculations, we identified elements that can be used to p-dope and n-dope TMD channels. Instead of directly creating defects in the TMD layer, our strategy relies on doping the gate-dielectric layer, which leads to modulation doping of the TMD channel. Using Ni as an example to dope HfO$_2$, we systematically showed that the carrier concentration in the TMD layer can be controlled by tuning the doping rate in HfO$_2$ as well as engineering the position of the defects relative to the channel layer. We expect that a similar investigation of the other identified candidate elements for p- and n-doping should yield similar results. While we performed explicit calculation only on the crystalline phase, we also expect the above identified elements to lead for doping of TMDs when doped into amorphous gate-oxides as long as they retain their local atomic coordination. Even though the primary focus of this paper is to control the carrier concentration in the TMD layer around the gate-dielectric, such an approach can certainly be extended to the other regions of the transistor (spacer, contact metal, etc.) where a larger hole doping concentration is not prohibitive of a smooth device operation.

\section{Methods}
We calculated the total energies using DFT calculations as implemented in Vienna \textit{ab~initio} simulation package (VASP)~\cite{Kresse96p11169}. The presented results are using the PBE functional. We verified that the results are qualitatively unaffected on including the effect of van der Waals interactions~\cite{Grimme06p1787, Hamada14p121103}. To compute the surface formation energies and to simulate the effect of the interface, a slab-vacuum model was used~\cite{Shin16p38}. The slab contained more than 9 bulk layers with a vacuum of more than 15~\AA. Prior studies have shown that the (111) surface is stable~\cite{Chen2008}. Therefore, the oxygen-terminated (111) surface was chosen in our study. A full structural relaxation was performed, keeping the three farthest HfO$_2$ sub-layers fixed to simulate the effect of bulk. Structural relaxation was done with a force convergence tolerance of 0.1 meV/{\AA} using a conjugate-gradient algorithm. The convergence criterion for the electronic self-consistent calculations was set to 10$^{-8}$~eV. A regular 8 $\times$ 8 $\times$ 1 $\Gamma$-centered $k$-point grid was used to sample the Brillouin zone with a plane-wave cutoff energy of 600~eV. When computing the effect of magnetism, we add a Hubbard $U$ as used in default pymatgen setting~\cite{Zhou04p235121}. For the calculations where the doping rate of Ni is varied a Hubbard $U$ of 4.5 eV was added to the Ni site to appropriately capture the effect of on-site Coulomb interaction~\cite{Dudarev98p1505, Tesch22p195153}.


\section{Acknowledgement}
The authors would like to thank Dr. M. Amachraa, Dr. Anub Halder, Dr. Ali Hamze, Dr. Jeong-J Cho, Dr. Yeonchoo Cho, Dr. Yoonhoo Ha, Dr. Minsu Seol, Dr. Eun-Kye Lee, and Dr. Kyung-Eun Byun for fruitful discussions.

\bibliography{citations}
\end{document}